\documentclass[12pt,preprint]{aastex}

\newcommand{\etal}{{et al.}}
\newcommand{\msun}{\thinspace M_\odot}  
  
\newcommand{\vect}[1]{\mbox{\boldmath$#1$}}

\newcommand{\rhoc}{\rho_{\rm c}}
\newcommand{\nc}{n_{\rm c}}

\newcommand{\ob}{\varepsilon_{\rm ob}}
\newcommand{\ar}{\varepsilon_{\rm ar}}

\newcommand{\cm  }{\,{\rm cm}^{-3} } 

\newcommand{\dfrac}[2]{{\displaystyle \frac{#1}{#2}}  }

\newcommand{\ap }{A_{\phi}}

\newcommand{\nf }{n_{\rm f}}
\newcommand{\omg}{\omega_{\rm c}}

\shorttitle{Formation of First-Star Binaries}
\shortauthors{Machida  \etal 2007}

\begin{document}
\title{Conditions for the Formation of First-Star Binaries}

\author{Masahiro N. Machida\altaffilmark{1} , Kazuyuki Omukai\altaffilmark{2}, Tomoaki Matsumoto\altaffilmark{3}, and Shu-ichiro Inutsuka\altaffilmark{1}} 

\altaffiltext{1}{Department of Physics, Graduate School of Science, Kyoto University, Sakyo-ku, Kyoto 606-8502, Japan; machidam@scphys.kyoto-u.ac.jp, inutsuka@tap.scphys.kyoto-u.ac.jp}
\altaffiltext{2}{National Astronomical Observatory of Japan, Mitaka, Tokyo 181-8588, Japan; omukai@th.nao.ac.jp}
\altaffiltext{3}{Faculty of Humanity and Environment, Hosei University, Fujimi, Chiyoda-ku, Tokyo 102-8160, Japan; matsu@i.hosei.ac.jp}

\begin{abstract}
The fragmentation process of primordial-gas cores during  prestellar 
collapse is studied using three-dimensional nested-grid hydrodynamics.
Starting from the initial central number density of
$\nc\sim10^3\cm$,
we follow the evolution of rotating spherical cores
up to the stellar density $\nc \simeq 10^{22}\cm$. 
An initial condition of the cores is specified by three parameters: 
the ratios of the rotation and thermal energies to the gravitational energy 
($\beta_0$, and $\alpha_0$, respectively), 
and the amplitude of the bar-mode density perturbation ($\ap$).
Cores with rotation $\beta_0 > 10^{-6}$ are found to fragment 
during the collapse. 
The fragmentation condition hardly depends on either the initial 
thermal energy $\alpha_0$ or amplitude of bar-mode perturbation 
$\ap$.
Since the critical rotation parameter for fragmentation 
is lower than that expected in the first star formation, 
binaries or multiples are also common for the first stars.
\end{abstract}

\section{Introduction}
More stars are in binaries or multiples than in singles 
in the solar vicinity
(Duquennoy \& Mayor 1992; Fischer \& Marcy 1992).
Is this also true of the first-generation stars?
Being very massive ($\ga 100M_{\sun}$; see e.g., Bromm \& Larson 2004), 
first-star binaries, if they are formed, can contribute greatly as 
progenitors of gamma-ray bursts (Bromm \& Loeb 2006;
but see Belczynski et al. 2006 for the opposite view) or 
gravitational wave sources (Belczynski \etal 2004).
However, binary formation in primordial gas has sometimes 
been considered unlikely, based on the fact that no 
cooling mechanism induces fragmentation of high-density 
($\ga 10^{4}{\rm cm^{-3}}$) cores (e.g., Ripamonti \& Abel 2004).

Rotation can change this picture.
Unlike the case of present-day star formation, 
where the angular momentum is effectively transported 
by magnetic braking \citep{basu94} or 
outflow \citep{tomisaka02}, in  first-star formation, 
the magnetic field is extremely weak \citep[e.g.,][]{takahashi05,ichiki06}
and thus expected to play little role in the angular momentum transport. 
With conserved angular momentum, the effect of rotation is 
enhanced in the course of the collapse of the core, and
a disk is expected to form.
Do such disks fragment to binaries or multiples? 
To answer this question, by three-dimensional hydrodynamics, 
Saigo \etal (2004) demonstrated that 
primordial-gas cores with rapid rotation indeed fragment into binaries.
Nonetheless, their analysis is not satisfactory in the following respects: 
First, they started calculation at rather high density 
$\sim 10^{10} {\rm cm^{-3}}$, and thus were forced to assume 
somewhat ad hoc initial conditions of high angular-momentum cores.
In practice, the angular momentum might have been transferred 
by the non-axisymmetric structure before reaching this density.
In addition, they presented only one model of a core experiencing 
fragmentation.
Recently, Clark, Glover, \& Klessen (2007) performed three-dimensional 
hydrodynamical simulation of star formation in a low-metallicity 
turbulent medium and found that rotation causes fragmentation even 
in metal-free gas, as suggested by Saigo et al. (2004). 
However, they do not discuss fragmentation conditions of the cores in their analysis.

In this paper, we calculated the evolution of primordial-gas rotating cores
for twenty orders of magnitude in the density contrast, 
starting from the formation of dense cores ($n\simeq 10^3\cm$) 
up to stellar-core formation ($n\simeq 10^{22}\cm$).
We adopt a realistic equation of state derived by Omukai et al. (2005),
rather than a simple polytropic relation $P\propto \rho^{1.1}$ 
used by Saigo et al. (2004). 
To determine the fragmentation conditions for the primordial-gas cores, 
we calculated 38 models with different values of rotation, thermal energy, 
and amplitude of non-axisymmetric perturbation.
Using the initial ratio of rotation to gravitational energies 
$\beta_0$, the conditions for a primordial-gas core to fragment are 
$\beta_0 > 10^{-6}$, while dependence of these conditions on either 
the thermal energy or the amplitude of 
non-axisymmetric perturbation is very weak.
These conditions are easily satisfied in the first-star forming cores, 
and thus binary formation is also important for first stars. 

The following is the plan of this paper.
In this paper, after first summarizing the basics of our model in \S 2, we describe the numerical method  in more detail in \S 3.
We present the results for the initially 
quasi-stable cores in \S 4, and those 
for highly gravitationally unstable cores in \S5.
The fragmentation criterion is examined in \S 6.
Finally in \S 7, summary and discussions are presented. 

\section{Model Settings}
\begin{figure}
\begin{center}
\includegraphics[width=150mm]{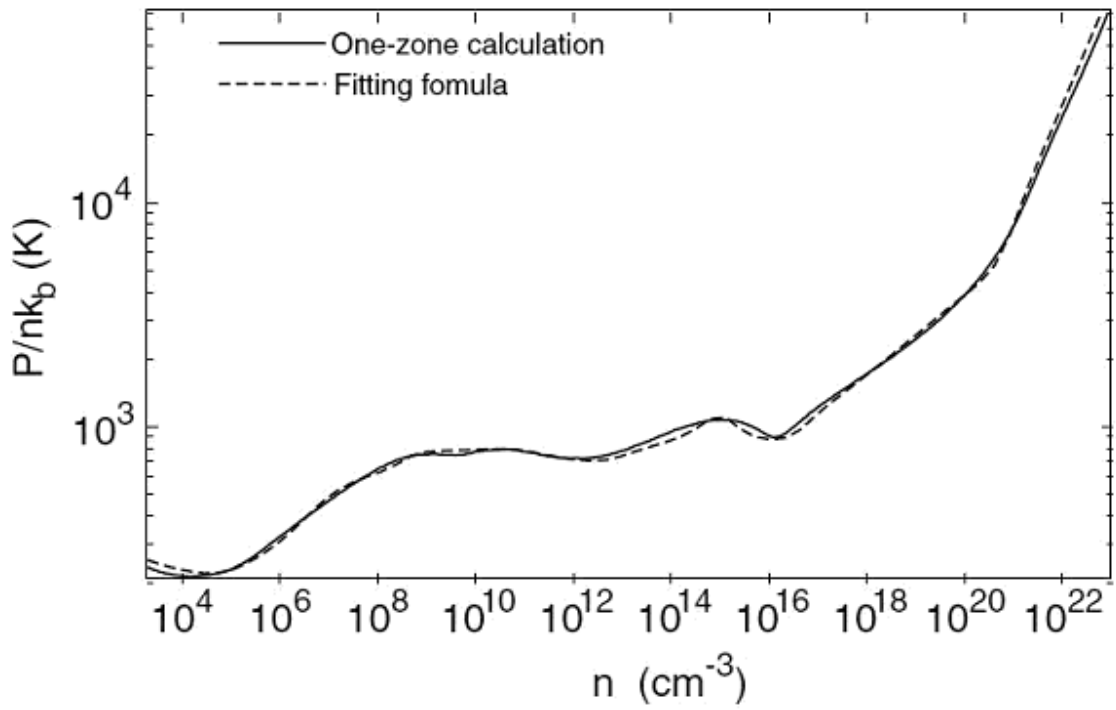}
\caption{
Gas pressure normalized by number density derived from a one-zone 
calculation (solid) and our fit used in the numerical simulation 
(dashed line).}
\label{fig:eos}
\end{center}
\end{figure}

We solve the equations of hydrodynamics including self-gravity:  
\begin{eqnarray} 
& \dfrac{\partial \rho}{\partial t}  + \nabla \cdot (\rho \vect{v}) = 0, & \\
& \rho \dfrac{\partial \vect{v}}{\partial t} 
    + \rho(\vect{v} \cdot \nabla)\vect{v} =
    - \nabla P -      \rho \nabla \phi, & 
\label{eq:eom} \\ 
& \nabla^2 \phi = 4 \pi G \rho, &
\end{eqnarray}
where $\rho$, $\vect{v}$, $P$, and $\phi$ denote the density, 
velocity, pressure, and gravitational potential, respectively. 
For gas pressure, we use a barotropic relation that approximates 
the result for a primordial-gas core by Omukai et al. (2005).
Omukai et al. (2005) studied the thermal evolution of low-metallicity 
star-forming cores using a simple one-zone model for the dynamics, 
where the core 
collapses approximately at the free-fall rate and the size of the core 
is about the Jeans length, as in the Larson-Penston self-similar
solution (Penston 1969; Larson 1969). 
On the other hand, detailed thermal and chemical processes in primordial 
gas are treated.
Their result and our fit to it are plotted against the number density
in Figure~\ref{fig:eos}.
To emphasize variations of pressure with density, 
we plot $P/n$, which is proportional to the gas temperature 
divided by the mean molecular weight.

\begin{figure}
\begin{center}
\includegraphics[width=140mm]{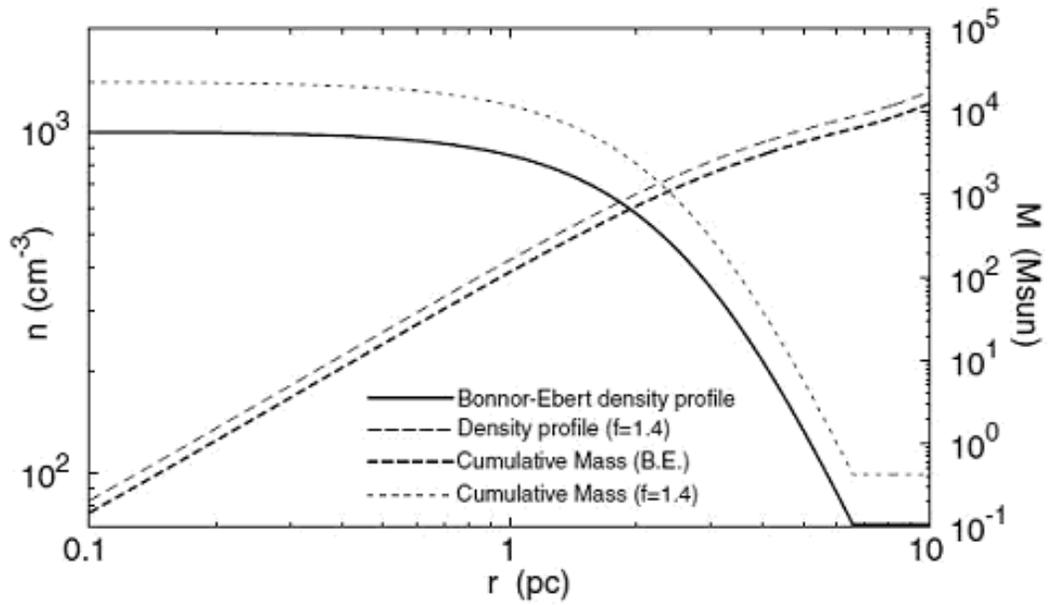}
\caption{The density ({\it thick solid line}) and enclosed mass ({\it thick dashed line}) distribution for the critical Bonnor-Ebert sphere 
as a function of the radius.
Those increased by a factor $f$=1.4 are also plotted by the thin lines.}
\label{fig:be}
\end{center}
\end{figure}

As an initial condition of the cores, we use a density profile 
increased by a factor of $f (\ga 1)$ from that of the critical 
Bonnor-Ebert (BE) sphere (Ebert 1955; Bonnor 1956).
For the critical BE sphere,  
the central number density is $n_{c,0} = 1 \times 10^3\cm$, 
and the initial temperature is about 250K at the center from 
our adopted barotropic relation (Fig.\ref{fig:eos}).
The critical radius of the BE sphere is  
$R_{\rm c} = 6.45\, c_s/\sqrt{4\pi G \rho_{BE}(0)} = 6.5$\,pc. 
Outside this radius, uniform gas density of 
$n_{\rm BE}(R_c)=71.1 f \cm$ is assumed.
The total mass contained in the critical BE sphere 
is $ M_{\rm c} = 6.2\times 10^3 \msun$, and thus our initial core is 
$f$ times more massive than this value. 
In Figure~\ref{fig:be}, the profiles of density and cumulative mass 
of the initial state are plotted. 

Initially the core rotates rigidly with angular velocity $\Omega_0$ 
around the $z$-axis.
The initial models are characterized by three non-dimensional parameters: 
the ratio of the thermal to gravitational energy ($\alpha_{0}$),
the ratio of the rotation to gravitational energy ($\beta_{0}$), 
and 
the amplitude of the non-axisymmetric perturbation ($A_{\phi}$). 
The models with different ($\alpha_{0}$, $\beta_{0}$) are constructed 
by changing the initial density enhancement factor ($f$) and 
angular velocity ($\Omega_0$).
Parameters $\alpha_{0}$ and $\beta_{0}$ are summarized 
in Table~\ref{table:init}, along with the density enhancement factor $f$.
Non-axisymmetric density perturbation of the $m=2$ mode, i.e., bar mode, 
is added to the initial cores. 
In summary, the density profile of the core is denoted as  
\begin{eqnarray}
\rho(r) = \left\{
\begin{array}{ll}
\rho_{\rm BE}(r) \, (1+\delta \rho)\,f & \mbox{for} \; \; r < R_{c}, \\
\rho_{\rm BE}(R_c)\, (1+\delta \rho)\,f & \mbox{for}\; \;  r \ge R_{c}, \\
\end{array}
\right. 
\end{eqnarray}
where $\rho_{\rm BE}(r)$ is the density distribution of the critical 
BE sphere, $\delta \rho$ is the non-axisymmetric density perturbation. 
For the bar($m=2$)-mode, 
\begin{equation}
\delta \rho = A_{\phi} (r/R_{\rm c})\, {\rm cos}\, 2\phi \; \; \; \mbox{for}\; \;  r \ge R_{c},
\end{equation}
where $A_{\phi}$ is the amplitude of the perturbation.
We use $A_{\phi}$ as a free parameter, and study cases with 
$A_{\phi} = 0.001-0.3$ (see Table~\ref{table:init}).
To disturb the $m=2$ symmetry in a high-density region 
\citep[for detail, see][]{matsu03}, we add the $m=3$ mode of velocity 
perturbation as
\begin{equation}
\Omega = \Omega_0 \left[ 1 + \Omega_3 (r/R_{\rm c})\, {\rm cos\, 3\phi} 
\right].
\end{equation}
The amplitude $\Omega_3$ is set to $10^{-5}$ in all the models.

\section{Numerical Method}
We adopt a nested grid method (for details, see Machida et al. 2005a; 2006a)
to obtain high spatial resolution near the center.
Each level of a rectangular grid has the same number of cells 
($ = 256 \times 256 \times 16 $), with the cell width $h(l)$ 
which is a function of grid level $l$.
The cell width is halved at each increment of the grid level.
The highest level of grids increases with time:
a new finer grid is generated whenever the minimum local Jeans 
length $\lambda _{\rm J}$ falls below $8\, h (l_{\rm max})$, 
where $h(l_{\rm max})$ is the cell width of the current finest grid. 
The maximum level of grids is restricted to $l_{\rm max} \le 30$. 
The generation of a new grid at the density maximum ensures 
the Jeans condition of \citet{truelove97} 
with a margin-of-safety factor of two.
We begin our calculations with four grid levels (i.e., $l_{\rm max}=4$).
The box size of the initial finest grid $l=4$ is chosen to be 
$2 R_{\rm c}$, where $R_{\rm c}$ is the radius of the critical BE sphere. 
The coarsest grid ($l=1$) thus has a box size of $2^4\, R_{\rm c}$. 
The mirror symmetry with respect to the $z$=0 is imposed.
At the boundary $r=2^4\, R_{\rm c}$, 
the ambient gas is set to rotate with angular velocity $\Omega_0$ 
(for more detail, see \citealt{matsu04}).

\section{Evolution of quasi-hydrostatic cores}
According to numerical simulations 
(Bromm et al. 1999; Abel et al. 2002; Yoshida et al. 2006),
in the case of first star formation, self-gravitating isolated 
cores form at the number density $10^3-10^4\cm$, where
temperature takes its minimum (see Fig.~\ref{fig:eos}).
These cores are close to the hydrostatic equilibrium, i.e., 
$\alpha_{0} \simeq 1$, which is a general property of cores 
forming by gravitational instability.
In this section, we study such cases as $\alpha_0 \simeq 1$, 
and we defer the effect of different $\alpha_0$ to Sec. 5.

In Figure~\ref{fig:final},
the final states of our simulation for the cores 
with $\alpha_0 \simeq 1$ are presented 
for various combinations of initial rotation parameter $\beta_0$ 
and bar-mode perturbation $\ap$.
The model parameters ($\alpha_0$, $\beta_0$, $\ap$) are summarized in 
Table~\ref{table:init} (models 1-24 for those in Fig.~\ref{fig:final}).
Also presented in Table~\ref{table:init} are the number density 
($n_{\rm f}$) at fragmentation, the number of  fragments, 
the separation between the fragments at the end of the calculation, 
and the mode of fragmentation (i.e., ring or bar-mode; see below).
Among models presented in Figure~\ref{fig:final}, for models 1-4 with 
the highest rotation parameter ($\beta_0 = 0.1$),
we need to adopt values of $\alpha_0$ ($=0.7$) smaller
than in the other models ($\alpha_0=0.83$) to initiate the collapse.
In Figure~\ref{fig:final} we see that cores with rotation 
$\beta_0 \ge 10^{-5}$ fragment for all values of 
bar-mode perturbation $\ap$ in our calculated range.

It is known that cores fragment only after they become either thin disks 
or elongated filaments (Tohline 1980).
Correspondingly, the mode of fragmentation can be classified into two types
(See Fig. \ref{fig:schematic}).
One is ring-mode fragmentation induced 
by rotation, and the other is bar-mode fragmentation 
induced by bar instability (Machida et al. 2004; 2005b).
The mode of fragmentation is determined by the axis ratio.
If the axis ratio does not grow sufficiently during the collapse,
the core becomes ring-like due to the rotation and then fragments into 
pieces (i.e., {\it ring-mode fragmentation}). 
On the other hand, a core with small rotation but with a
large enough bar-mode perturbation results in an elongated structure, and 
fragments due to the bar instability (i.e., {\it bar-mode fragmentation}).
For example, models in the second column of Figure~\ref{fig:final} 
experience ring-mode fragmentation, while in the fourth column where 
$\beta_0 \ge 10^{-3}$ the cores fragment through the bar mode.
The mode of fragmentation determines the fate of the fragments 
to some extent.
Survival of fragments and formation of binary/multiple
are easier in the ring mode due to large orbital angular momentum
between the fragments, while in the bar mode, they tend to merge, and fragmentation is repeated in the merged core \citep{matsu03,machida05b}.
 
Our calculations are terminated when either the Jeans condition 
is violated in the maximum ($l_{\rm max}=30$) grid level or 
the maximum density exceeds $\nc \gtrsim 10^{22}\cm$.
Since the former situation easily occurs when the fragments leave the center, where the grid is finest, 
after the fragmentation we can continue calculation only for a short duration.
Due to this limitation, our calculation covers only a rather low-density 
range for models with high $\beta_0$ (models 1-4), where fragmentation 
occurs at low densities ($\nc \simeq 10^{11}\cm$).
In these models, the cores might experience another episode of 
fragmentation if we continued calculation further. 
For the other models in Figure~\ref{fig:final}, 
our calculation covers the entire evolution until the stellar core 
formation after the equation of state becomes adiabatic.

\subsection{Case Studies}
In this section, we see the evolution of cores 
in models 22, 6 and 8, which are 
examples of a non-fragmenting core, a core fragmenting in the ring-mode, 
and in the bar-mode, respectively.  

We define here the following three quantities which describe 
the state of cores at a given time for later use.
To examine the evolution of rotation velocity around the center, 
we use angular velocity normalized by the free-fall timescale; 
$\omega_{\rm c} \equiv \Omega_{\rm c}/(4\pi G \rhoc)^{1/2}$.
Hereafter, we call this quantity the normalized angular velocity.
For a rigidly rotating sphere with uniform density,  
the normalized angular velocity is related 
to the ratio of the rotation to the gravitational energy  
($\beta_{\rm c}$) by 
\begin{equation}
\beta_{\rm c} = \dfrac{\Omega_{\rm c}^2 R^3}{3 G M} 
= \dfrac{\Omega_{\rm c}^2}{4\pi G \rhoc},
\label{eq:beta}
\end{equation}
where $R$ and $M$ are the radius and mass of the sphere, respectively.
Even when the core deviates somewhat from the sphere 
in the course of the collapse, equation~(\ref{eq:beta}) remains 
a good indicator for rotation (Machida et al. 2005a; 2006a).
Also, to see the degree of deformation from a sphere and approach 
to a disk, we define oblateness of the core as
$ \ob \, \equiv \, (h _l h _s) ^{1/2} / h _z $, 
where $ h _l $, $ h _s $, and $ h _z$ are, respectively, 
lengths of the major, minor, and $z$-axes derived from 
the moment of inertia for the high-density gas 
$ \rho \, \ge \, 0.1  \rho _{\rm c} $, 
i.e., density higher than 10\% of the central value. 
Also, as the degree of deviation from a circle when projected 
on the $x-y$ plane, the axis ratio is defined as 
$ \ar \, \equiv \, h _l/ h _s - 1$.

\subsubsection{Non-fragmenting Core: Model 22}

As an example of cores that do not experience fragmentation, 
we describe the evolution in model 22, whose parameters are 
($\alpha_{0}$, $\beta_0$, $\ap$) = (0.83, $10^{-6}$, $10^{-2}$).
In this case, a hydrostatic core (``stellar core'') forms without 
fragmentation. 
This structure is shown in several evolutionary stages in Figure~\ref{fig:model22}.
Panel {\it a} shows the initial state.
We adopt the critical BE sphere as an initial state (Sec.2),
but add $1\%$ of density perturbation for triggering gravitational 
collapse.
The initial density contrast is thus 14 between the center and 
the ambient medium.

Since the rotation is very slow ($\beta_0=10^{-6}$) in this model,
the collapse becomes almost spherically symmetric in the 
early phase, as seen in panels {\it b-d}.
In fact, the infall motion is almost radial, i.e., the radial component 
of the gas velocity is much larger than the azimuthal component, 
$v_{\rm r}\gg v_{\rm \phi}$.
The effect of rotation becomes important with the collapse, 
and deviation from the spherical collapse 
becomes obvious around $\nc \sim 10^{17}\cm$
(see Figs.~\ref{fig:model22}{\it e}, where $\nc = 7.1\times 10^{18}\cm$).
Now the azimuthal velocity is comparable to the radial one 
($v_{\phi} \simeq v_{\rm r}$), and the gas falls to the center 
with significant rotation.

In Figure~\ref{fig:5a}, we plot the normalized angular velocity 
(Fig.~\ref{fig:5a}{\it a}), oblateness (Fig.~\ref{fig:5a}{\it b}), and 
axis ratio (Fig.~\ref{fig:5a}{\it c}) against the central number density 
for this case (model 22) along with those with 
different $\beta_0$ but the same $\ap=10^{-2}$ (models 2, 6, 18) 
for comparison.
Those quantities are calculated for the central region whose density 
is higher than 10\% of the central value.

For the spherical rigid-body rotation, 
angular velocity increases as $\Omega_{\rm c}\propto \nc^{2/3}$ 
owing to the conservation of angular momentum, 
and thus normalized angular velocity $\omg \propto \nc^{1/6}$.
In model 22, this spherical collapse phase continues 
until $\nc \sim 10^{19}\cm$, where the saturation 
of the normalized angular momentum around 
$\omega_{\rm c} \simeq 0.2-0.3$ is observed. 
This indicates that a thin disk begins to form at the center, 
for which the normalized angular velocity increases
as $\omega_{\rm c} \propto \nc^{\gamma-1 /2}$, 
which is almost constant in our case where $\gamma \simeq 1.1$. 
This can be derived from the conservation of specific angular momentum 
$j/M = \pi \varpi^2 \Omega_{\rm c} /\pi \varpi^2 \Sigma =const. $, 
where $\varpi$ is the radius of the disk and 
the column density $\Sigma$ can be expressed as 
$\Sigma \simeq \rho H$ using the disk scale height
$H \simeq c_{s}/\sqrt{G \rho} \propto \rho^{\gamma-1/2}$, 
where $c_{s}$ is the sound speed.

The formation of the disk can be seen in the behavior of the oblateness.
For $\nc \la 10^{19}\cm$, the collapse is spherical and 
the oblateness remains close to unity (see Fig.~\ref{fig:5a}{\it b}). 
After the normalized angular momentum reaches the saturation level and 
the radial collapse is hindered to some extent by centrifugal force, 
the oblateness begins to increase.
When the central density reaches $\nc = 3 \times 10^{21} \cm$, a stellar core with an initial mass of $6.7 \times 10^{-3} \msun$, forms at the center, surrounded by a shock.
Figure~\ref{fig:model22}{\it f} shows the structure immediately after 
the stellar core formation.
In this figure, the surface of the stellar core is indicated 
by the thick black line.
The density and mass of the stellar core at the formation in our model 
are well in agreement with the results of Omukai \& Nishi (1998), 
who calculated the collapse of a primordial star-forming core  
including detailed radiative and chemical processes 
under the assumption of spherical symmetry.
After the formation, the stellar core becomes round in shape and
the oblateness stops increasing.
The oblateness can grow remarkably only
after the saturation of $\omega_{\rm c}$ and before the formation 
of the stellar core.
In this model, the saturation of $\omega_{\rm c}$ occurs only slightly 
before the stellar core formation, owing to a small initial rotation $\beta_{0}$.
Thus, the core does not become oblate enough to 
fragment in the ring mode (see Sec 4.1.2). 

Next, we see the growth of the bar-mode perturbation $\ar$
in Figure~\ref{fig:5a}{\it c}.
Recall that we add 1\% of bar-mode density perturbation 
to the initial state, i.e., $\ap=10^{-2}$.
The axis ratio of the core in model 22 remains very small 
throughout the evolution ($\ar \lesssim 10^{-2}$).
The growth of the bar-mode perturbation can be described by 
a linear perturbation theory developed by
\citet{hanawa00} and \citet{lai00} for small deviation 
from the spherical self-similar solution.
They showed that a collapsing spherical core with a polytropic
index $\gamma$ is unstable against a bar-mode perturbation when $\gamma \le
1.097$, and the perturbation of the bar mode grows as $\ar \propto
\nc^s$, with $s$ being a constant for a given $\gamma$. 
For the isothermal case ($\gamma = 1$), for example, $s=0.177\,(\simeq 1/6)$: 
the growth rate is the same as that of the normalized angular velocity 
$\omega_{\rm c}$ \citep{hanawa99}.
For more reference, see Figure~14 of \citet{omukai05}, where the power 
index $s$ is shown as a function of $\gamma$. 
In our case, the polytropic index ranges in $1\lesssim \gamma \lesssim 1.15$ 
(see Fig.~\ref{fig:eos}), corresponding to the range of the power index
$-0.4 \lesssim s \lesssim 0.2$.  
For $10^8\cm \lesssim \nc \lesssim 10^{16}\cm$, $\gamma$ becomes 
$\simeq 1$, and then $\ar$ grows (Fig.~\ref{fig:5a}{\it c}),
although the growth rate is somewhat lower than in the isothermal case.

We show the evolution of the radial density distribution 
averaged around the rotation axis in Figure~\ref{fig:6a}{\it a}.
With only slow rotation, the evolution of both density and 
velocity distributions resembles closely a result of the 
spherically symmetric calculation (Omukai \& Nishi 1998). 
The collapse proceeds in a self-similar way with a polytrope 
index $\gamma \simeq 1.1$:
the evolution of density distribution resembles closely the 
Larson-Penston self-similar solution, whose density gradient 
in the envelope is $\propto r^{-2/(2-\gamma)} \simeq r^{-2.2}$.
The appearance of a shock front due to the stellar core formation 
is seen in the velocity distribution at state 6 
(Figure~\ref{fig:6a}{\it b}). 

\subsubsection{Ring-mode Fragmentation: Model 6}
Here, we see an example of a core with rapid rotation.
In Figure~\ref{fig:7}, we show the evolutionary sequence of model 6, 
whose parameters are ($\alpha_0$, $\beta_0$, $\ap$) 
= (0.83, $10^{-2}$, $10^{-2}$).
Rotation energy of this model is $10^4$ times larger than 
that of model 22 above, 
while the other parameters $\ap$ and $\alpha_0$ are the same.
In contrast to model 22, where the collapse is almost spherical, 
here rotation plays a major role.
After some contraction, the core becomes ring-like, owing to the centrifugal
support, and eventually fragments. 

The core is still spherical early in the evolution 
(Fig.~\ref{fig:7}{\it a}; $\ob = 1.2$, $\nc = 3 \times 10^7\cm$).
With further contraction, rotation makes it oblate perpendicular 
to the $z$-axis
(Fig.~\ref{fig:7}{\it b}; $\ob = 3.8$, $\nc= 2 \times 10^{11}\cm$). 
As in model 22 above, the normalized angular velocity 
grows initially as $\omg \propto \nc^{1/6}$ and then 
saturates around $\omg \simeq 0.2-0.3$ (Fig.~\ref{fig:5a}{\it a}).
In this model, however, the saturation occurs earlier 
at $\nc \simeq 10^7\cm$ than in model 22 ($\simeq 10^{19}\cm$), 
owing to faster initial rotation. 
The normalized angular velocity oscillates 
around the saturation value $\omg \simeq 0.2-0.3$ thereafter,
which corresponds to $\beta_{\rm c}=0.04-0.09$ 
by the relation (\ref{eq:beta}).
In other words, after the ratio $\beta_{\rm c}$ reaches a certain level of 
$0.04-0.09$, the core contracts, keeping a constant 
ratio between the centrifugal and gravitational forces near the center.
The approach to a disk-like configuration can be observed in 
the increase of oblateness 
(the dashed line in Fig.~\ref{fig:5a}{\it b})
after the saturation of the normalized angular velocity.
The oblateness also saturates at $\ob \simeq 10$ and then oscillates 
after that.
Note that the face-on view on the $x-y$ plane remains quite circular
even after the thin disk formation.
For example, the axis ratio $\ar$ is $\sim  10^{-3}$ 
at the time of Panels {\it a} and {\it b} of Figure~\ref{fig:7}.

As can be seen in the middle and lower panels of Figure~\ref{fig:7}{\it c},
two disks are nested, and each disk is sandwiched by horizontal shocks above and below. 
The outer and inner shocks exist at $z=\pm 0.2$\,AU and $z=\pm 0.05$\,AU, 
respectively.
Outside the outer disk, cavities exist along the $z$-axis.
The emergence of the shocks is related to the oscillatory behavior of
oblateness after its saturation:
a pair of shocks appear whenever the core becomes maximally oblate and 
then bounces back \citep{matsumoto97,machida05a,machida06a}.
In this model, two maxima of the oblateness appear around 
$\nc \simeq 10^{13}\cm$ and $10^{17}\cm$ (Fig.~\ref{fig:5a}{\it b}).
Correspondingly, the outer shock appears at $\nc \simeq 10^{13}\cm$ 
and the inner shock at $\nc \simeq 10^{17}\cm$.

The axis ratio of an oblate core grows faster than the linear 
theory predicts (see Fig.~\ref{fig:5a}{\it c}).
Although the core deforms to an ellipsoid by this effect
(Fig.~\ref{fig:7}{\it c}; $\nc \simeq 5\times 10^{18}\cm$),
the deformation is not sufficient owing to the initial 
small value of $\ap$.
Consequently, the core eventually transforms to a ring rather 
than to an ellipsoid, and fragments at  
$\nc \simeq 7\times 10^{18}\cm$ into two pieces with 
almost equal mass of $\simeq 1 \times 10^{-2}\msun$ 
(Fig.~\ref{fig:7}{\it d}).
The rapid drop of the normalized angular velocity 
at $\nc \simeq 10^{18}\cm$ 
(dashed line of Fig.~\ref{fig:5a}{\it a}) can be attributed to 
a rapid increase in the oblateness of the core 
(i.e., deformation to a flat disk shape),
and consequent redistribution of the angular momentum of the core
due to this non-axisymmetric density distribution.

When the density reaches $n \simeq 10^{20}\cm$, 
a stellar core forms in each fragment. 
Each core is surrounded by a shock
(Fig.~\ref{fig:7}{\it e} and {\it f}).
Although the stellar cores are initially ellipsoidal in shape 
(Fig.\ref{fig:7}{\it e}), they subsequently become spherical  
(Fig.~\ref{fig:7}{\it f}).
The mass of each stellar core reaches $3.1 \times 10^{-3}\msun$ each
at the end of the calculation.
The two cores continue to part, and the final separation 
is $R_{\rm sep} = 0.36$\,AU.
Although the main bodies of the cores approach spherical in shape after 
the formation, their envelopes become more elongated, like tails. 
They eventually become detached (see the upper panel of Fig.~\ref{fig:7}{\it f}).
Having density peaks inside [at $(x, y) \simeq (\pm0.17, \pm0.24)$], 
these ``tails'' may hatch other stellar cores.

The evolutionary sequences of number density and velocity 
before fragmentation are shown in Figures~\ref{fig:6b}{\it a} and 
{\it b}, respectively.
Despite the deformation to a disk-like structure,
the evolution of the core is well described by the Larson-Penston 
self-similar solution until the fragmentation \citep[c.f.,][]{saigo00}.
Similar behavior in the isothermal collapse has been found by  
\citet{matsumoto97}.

\subsubsection{Bar-mode Fragmentation: Model 8}

Here, we see model 8, whose parameters are
($\alpha_0$, $\beta_0$, $\ap$) = (0.83, $10^{-2}$, $0.3$). 
The initial core has non-axisymmetric perturbation 30 times larger  
than in model 6, while the other parameters are the same. 
The evolutionary sequence of this model is shown in Figure~\ref{fig:8}.
A non-axisymmetric structure, or a ``bar'', already appears  
in an early phase. 
The bar becomes more elongated with collapse and eventually fragments.
Evolution of the normalized angular velocity ($\omg$), oblateness ($\ob$), 
and axis ratio ($\ar$) are plotted in Figures~\ref{fig:5b}{\it a}, 
{\it b}, and {\it c} (dashed lines).

Like in the previous examples, the central portion evolves keeping 
a spherical shape until disk formation (Fig.~\ref{fig:8}{\it a}; 
$\nc = 4.7\times 10^8\cm$).
Evolution of both the normalized angular velocity and the 
oblateness is also very similar to those of previous models 
(Fig.~\ref{fig:5a}{\it a}, {\it b}).
The oblateness saturates around $\ob \simeq 7$ at $\nc \simeq 10^{13}\cm$ 
and it subsequently oscillates (dashed lines of Fig.~\ref{fig:5b}{\it b}).
When the oblateness reaches its maxima, a thin disk bounded 
by shocks forms, as in model 6 (middle and lower panels
of Fig.~\ref{fig:8}{\it c}).
However, owing to its high initial value, 
the non-axisymmetric perturbation has grown significantly 
by the time of the disk formation, i.e., the saturation of $\omega_{c}$.
The growth of the axis ratio is accelerated
even more after the disk formation and a clear bar-like structure develops 
(Fig.~\ref{fig:8}{\it d}). 
The bar elongates further and eventually fragments at 
$\nc = 1.9 \times 10^{19}\cm$.
By this moment, the axis ratio reaches $\ar=31.4$.
This value is similar to the axis ratio at the fragmentation 
found by Tsuribe \& Omukai (2006), whose calculation is limited to 
non-rotating cases. 
Inside the fragments, the stellar cores form 
after the equation of state becomes 
adiabatic (Fig.~\ref{fig:8}{\it e, f}).

The separation between the fragments is $R_{\rm sep}=0.4$\,AU and 
the mass of a fragment is $3 \times 10^{-2}\msun$ at the end of 
the calculation.
The fragmentation epoch and separation are similar to those in model 6, 
which implies that the initial amplitude of the non-axisymmetric 
perturbation does not affect these quantities. 
In contract to the case of model 6,
the fragments approach each other after 
fragmentation as a result of their small orbital angular momentum.
Judging from the final state of our calculation, merger of the 
fragments appears inevitable although it has not yet occurred 
within our run.

\subsection{Effect of rotation parameter $\beta_{0}$}

In this section, we summarize effects of different initial rotation 
parameters $\beta_0$ on the evolution.
As examples, in the second column of Figure~\ref{fig:final}, we compare six models 2, 6, 10, 14, 18 and 22, which have
different $\beta_0$ ($=10^{-6}-0.1$) but with the same $\ap$($=10^{-2}$).
The normalized angular velocities, oblatenesses, 
and axis ratios are plotted in Figures~\ref{fig:5a}{\it a}, \ref{fig:5a}{\it b},
\ref{fig:5a}{\it c} for four of them (models 2, 6, 18 and 22). 

Except for the fastest rotation case of model 2, 
normalized angular velocities increase in the low density regime 
as $\omega_c \propto \nc^{1/6}$, 
which is the spherical collapse relation.
The normalized angular velocity saturates 
around $\omega_{\rm c} \simeq 0.2-0.3$ in all cases.
Note that the saturation values of $\omega_{\rm c}$ are similar in all models,
although the epoch of saturation is earlier (i.e., at lower densities) 
for those with larger initial rotation.
After the saturation, the normalized angular velocity oscillates 
around the saturation value $\omg \simeq 0.2-0.3$ until fragmentation.
Similar behaviors of the angular velocity have been found 
by \citet{matsu03} and \citet{machida06a}, who showed that the normalized 
angular velocity converges to a certain value in the self-similar collapse.
In the case with the fastest rotation (model 2; $\beta_0 = 0.1$),  
since the saturation level $\omg \simeq 0.2-0.3$ 
($\beta \simeq 0.04-0.09$) has been reached 
from the onset, the normalized angular velocity remains almost constant 
until fragmentation.

The shape of the cores remains almost spherical, namely, the oblateness
is small $\ob \simeq 1$, until the saturation of the normalized angular 
velocity.
It is only after the saturation of $\omega_{\rm c}$ that the oblateness 
begins to increase remarkably (Fig.~\ref{fig:5a}{\it a, b}).
In the largest rotation case of model 2,
the oblateness can increase from the onset since the saturation 
level of $\omega_{\rm c}$ has already been attained already in the initial state. 
Like $\omega_{\rm c}$, the oblateness also saturates at $\ob \simeq 10$
and oscillates thereafter around this value until fragmentation.

From this behavior of oblateness, we conclude that 
the saturation of $\omega_{\rm c}$ to $\simeq 0.2-0.3$ is necessary 
for the formation of a thin ($\ob \simeq 10$) disk.
A thin disk can continue collapsing with conserved angular momentum 
for effective adiabatic index $\gamma \leq 1$ \citep{saigo00}.
For $\gamma > 1$, the central portion of the disk 
eventually supported by rotation and its collapse is halted, 
leading to fragmentation of the disk. 
In our case, where $\gamma >1$ in most of the density range, 
such a thin disk eventually fragments. 
Thus {\it the saturation of $\omega_{\rm c}$ is a prerequisite for 
fragmentation of the core}.

The evolution of the axis ratio is presented in Figure~\ref{fig:5a}{\it c}.
Also shown by a straight line is that for the isothermal gas in the 
linear theory $\ar \propto \nc^{1/6}$ (see Sec. 4.1.1).
Before the formation of thin disks,
the axis ratios decrease for $\nc \la 10^8\cm$ 
and then increase in higher densities by obeying the linear 
theory.
After the flat disk with $\ob>7-10$ is formed, 
the growth rate of the axis ratio is more enhanced than 
the linear theory predicts.
In fact, the axis ratio continues to increase even when
the polytropic index $\gamma$ is larger than 1.1
for $\nc > 10^{16}\cm$.
A similar effect is known in the isothermal collapse:
a collapsing disk exhibits a larger growth rate of the
bar-mode perturbation than a sphere \citep{nakamura97, matsu99}.

Since fragmentation occurs only after the disk formation,
cores with larger initial rotation tend to fragment 
at lower densities, and the separations between fragments are wider.
For example, in model 2 ($\beta_0=0.1$), with the fastest rotation, 
the fragmentation density is $2 \times10^{11}\cm$ 
and the separation between the fragments is $R_{\rm sep}=600$ AU 
at the end of the calculation, while in model 18 ($\beta_0 = 10^{-5}$), 
they are $6 \times 10^{20} \cm$ and 0.09AU, respectively. 
For slower rotation ($\beta_0=10^{-6}$; model 22), 
fragmentation does not take place.
We also calculated a few cases with even smaller rotation energies 
and confirmed that the cores do not fragment.

\subsection{Effect of bar-mode perturbations}
Next, in order to study the effect of initial bar-mode perturbation $\ap$, 
we focus on four models (models 5-8) 
with the same $\beta_0$ ($=10^{-2}$) 
but with different $\ap$ 
($=0.001$, 0.01, 0.1, and 0.2 for models 5, 6, 7, and 8, 
respectively).
Their final states are presented in the second row 
of Figure~\ref{fig:final}.

With increasing $\ap$, the mode of fragmentation changes from 
the ring to bar mode.
In models with small $\ap$ ($\le 0.01$; models 5 and 6), 
a ring forms near the center owing to centrifugal force
and subsequently fragments into several pieces.
On the other hand, for larger $\ap$ ($\ge 0.1$; models 7 and 8),
the central region transforms from a disk 
to a bar-like structure before fragmentation. 
In model 7, the core takes a ring-like structure near the center 
whereas a bar-like one in a larger scale.
The model with largest $\ap$ (Model 8) clearly takes a bar-like structure.

Whether fragmentation occurs or not appears to be solely determined 
by the rotation parameter $\beta_0$.
Cores with $\beta_0 > 10^{-6}$ always experience fragmentation 
before the formation of stellar cores.
In addition, fragmentation epoch and separation between fragments depend only weakly on bar-mode perturbation $\ap$.
The fragmentation density $\nf$ (i.e., epoch) ranges within an 
order of magnitude despite more than two orders of magnitude 
difference in $\ap$: $\nf=4 \times 10^{18}\cm$ ($\ap=0.001$; model 5), 
$7 \times 10^{18}\cm$ ($0.01$; model 6), 
$7 \times 10^{18}\cm$ (0.1; model 7), 
and $2 \times 10^{19}\cm$ (0.2; model 8).
We can see a weak trend that for larger $\ap$, 
the fragmentation densities are higher. 
The larger bar-mode perturbation $\ap$ extracts more angular momentum 
from the central region, thereby delaying formation of 
a quasi-rotation-supported disk and thus fragmentation.

The evolution of normalized angular velocity $\omega_{c}$, 
oblateness $\ob$, and 
axis ratio $\ar$ are shown in Figures~\ref{fig:5b}{\it a, b, c} for 
models 5, 6, 7, and 8.
With the same $\beta_0$, evolutionary tracks for both $\omega_{c}$ 
and $\ob$ are very similar for all these models
until $\nc \simeq 10^{15}\cm$ (Fig.~\ref{fig:5b}{\it a, b}).
For higher but still below fragmentation densities 
($\nc\simeq 10^{15}-10^{18}\cm$),
these tracks deviate from each other slightly 
owing to different shapes of the cores around the center 
(Fig.~\ref{fig:final}): 
disks remain highly circular in low $\ap$ models (5, 6 and 7), 
while an elongated bar forms in model 8 with high $\ap$.
As seen in Figure~\ref{fig:5b}{\it c},
all the curves for $\ar$ have the same shape and are only
shifted  upward or downward in relation to one another.
That is, their growth rates are common but initial values 
are different.
Fragmentation occurs via the ring mode in models 5-7
while via the bar-mode in model 8. 
At the fragmentation, the axis ratios are less than unity for 
models 5-7 ($\ar = 4.2 \times 10^{-2}$, 0.47, and 0.46 
for models 5, 6, and 7, respectively), 
whereas it reaches as high as 31.4 for model 8.

In summary, the initial amplitude of bar-mode perturbation 
does affect the fragmentation mode, but the fragmentation epoch and 
separation of fragments are mostly determined 
by the rotation parameter $\beta_0$.

\section{Evolution of Highly Unstable Cores:
\\ Effect of initial thermal energy $\alpha_0$}

In the previous section, we investigated the evolution of cores with 
thermal energy comparable to gravitational energy ($\alpha_0 \approx 1$). 
Although this setting is probably most realistic for initial cores
according to numerical simulations of gravitational fragmentation 
(e.g., Bromm et al. 2002; Yoshida et al. 2006),
to see dependence of core evolution on initial thermal energy 
parameter $\alpha_0$,
we study cases with $\alpha_0 \ll 1$ in this section.

Figure~\ref{fig:9} shows the core structure at the end of the calculation
for models 25-28.
These models have the same initial rotation parameters ($\beta_{0}=10^{-3}$) 
and bar-mode perturbations ($\ap=0.01$), 
but different thermal energies ($\alpha_0$).
Figure~\ref{fig:10} shows the evolution of normalized angular 
velocity $\omg$, oblateness $\ob$, and axis ratio $\ar$.
The difference in $\alpha_{0}$ hardly affects the evolution of 
$\omega_{\rm c}$ and $\ob$, although, for models with smaller $\alpha_0$, 
both the $\omega_{\rm c}$ and $\ob$ saturate at slightly higher densities 
and at slightly higher values (Fig.~\ref{fig:10}{\it a, b}).
On the other hand, the axis ratios $\ar$ evolve quite differently among 
these models, in particular, in low densities $\nc \lesssim 10^{10}\cm$
(Fig.~\ref{fig:10}{\it c}):
$\ar$ increases more for the lower $\alpha_{0}$ cores.
For higher densities, where the evolution of the cores already 
converge to the self-similar solution, the growth rate of $\ar$
becomes the same for all the models, but this convergence occurs 
later for the lower $\alpha_{0}$ cores. 
Therefore, cores with smaller $\alpha_{0}$ tend to have larger $\ar$, 
and then to fragment in the bar mode. 
For example, in models 25 and 26 with small $\alpha_{0}$ (=0.2, 0.4),  
the bar-mode perturbation grows sufficiently and the cores
fragment in the bar mode (Fig.~\ref{fig:9}{\it a, b}), while
in models 27 and 28 ($\alpha_{0}=0.6$ and 0.8), fragmentation occurs 
in the ring mode (Fig.~\ref{fig:9}{\it c, d}). 

\section{Fragmentation Condition}
As seen in Figure \ref{fig:final}, 
the condition for fragmentation appears to be
\begin{equation}
\beta_0 > \beta_{\rm crit} \sim 10^{-6}-10^{-5},   
\label{eq:beta0}
\end{equation}
i.e., fragmentation occurs if the initial core has rotation 
energy $10^{-6}$-$10^{-5}$ times the gravitational energy.
Or, in terms of angular velocity of the initial core, 
this can be rewritten as
\begin{equation}
\Omega_0 > 3.67 \times 10^{-17} \left( \dfrac{n_0}{10^3 \cm} \right)^{2/3} 
\  {\rm s^{-1}},
\label{eq:omg}
\end{equation}
where $\Omega_0$, and $n_0$ are the angular velocity 
and the central number density of the initial core, respectively,  
and we set $\beta_{\rm crit}=10^{-6}$.
The dependence of equation~(\ref{eq:omg}) on the density comes 
from the fact that the angular velocity increases 
$\Omega \propto \nc^{2/3}$ in the case of a spherically collapsing core.

Next, we present a physical explanation for the fragmentation condition 
(\ref{eq:beta0}).
As in Sec.4.2 and 4.3, the formation of a thin disk with oblateness 
$\ob \simeq 10$ is necessary for fragmentation.
This must be attained before the equation of state becomes adiabatic 
at $\nc \simeq 10^{20}\cm$ since the core becomes spherical and 
fragmentation is prohibited thereafter.
The oblateness begins to grow remarkably only after the normalized 
angular velocity 
becomes close to the saturation value $\simeq 0.2-0.3$, 
and the level of $\ob \simeq 10$ is reached after further contraction 
of about four or five orders of magnitude in density.
Therefore, as a condition for fragmentation, 
the normalized angular velocity must 
reach the saturation value ($\omega_{c} \simeq 0.2-0.3$, or 
$\beta_{\rm c} \simeq 0.1$) before $\sim 10^{15}-10^{16}\cm$.
By using the relation 
$\beta_{\rm c}=\sqrt{\omega_{\rm c}} \propto \nc^{1/3}$ 
for the spherical collapse, 
this can be translated to that on initial 
(i.e., at $\simeq 10^{3} {\rm cm^{-3}}$) rotation parameter as 
$\beta_{0} > (0.5-1)\times 10^{-5}$.
This condition coincides well with our empirical relation 
(\ref{eq:beta0}).

In deriving the above condition, we neglect angular momentum 
transfer before the formation of a thin disk.
As seen in Sec. 4.3, with the higher the bar-mode perturbation $\ap$, 
the more angular momentum is transported. 
Thus, cores with higher initial $\ap$ would 
need higher $\beta_{0}$ for fragmentation to compensate the angular momentum transfer.
This effect appears to be modest and is not observed clearly 
in our calculation.

\section{Summary and Discussion}

We have studied the gravitational collapse of 
zero-metallicity star-forming cores for different 
initial conditions, which are specified by 
the following three parameters, i.e., 
thermal energy, rotation energy, and 
amplitude of non-axisymmetric perturbations.
Since the thermal energy of star-forming cores 
is expected to be similar to the gravitational energy, 
particular attention is paid to such cores.
Initial conditions of the cores where 
fragmentation occurs are surveyed.
We found that the cores with initial rotation energy
higher than $10^{-6}-10^{-5}$ times the gravitational 
energy fragment into a few pieces.
Dependence on the other parameters is weak and 
is not observed clearly in our calculation.
The physical explanation for this fragmentation condition 
is also presented.

How much angular momentum do initial cores have in reality?
According to cosmological simulations by \citet{yoshida06},
first-star forming cores have the rotation parameter 
$\beta\simeq 0.1$.
In discussing the fragmentation condition,
the rotation parameter near the center should be used rather 
than that averaged over the core.
Matter in the outer part tends to have larger rotation 
energy than gravitational energy, the value of the rotation 
parameter $\beta$ becomes larger if outer material is 
included in calculations.
Even though the central value of the rotation parameter 
might be significantly smaller than 0.1 owing to this effect,
it is probably higher than our threshold value 
for fragmentation ($10^{-5}-10^{-6}$) with a wide margin.
Therefore, most of the cores are expected to fragment and proto-binaries 
are formed even in primordial-gas clouds.

In Figure~\ref{fig:12}, we plot the separations between fragments 
$R_{\rm sep}$ at the end of the calculation 
against the fragmentation epoch $n_{\rm f}$
for all the models where fragmentation is observed.
The large crosses show ranges of fragmentation epochs (horizontal) and 
fragmentation scales (vertical) for models with $\beta_0 >0.02$ (dotted line) 
($\beta_0<0.02$, solid line, respectively).
The Jeans length (left axis) and gas temperature (right axis) are 
also shown.
We can see that the separation between fragments is 
approximately equal to the Jeans length at the fragmentation.
Another striking feature is that fragmentation densities are clustered 
in two epochs of $n \approx 10^{12}\cm$ and $10^{21}\cm$.
Cores with rapid rotation ($\beta_0 > 0.02$; symbols '$+$') 
fragment in the earlier epoch ($\nc \simeq 10^{10}-10^{14}\cm$) 
with larger separation $\simeq 10-1000$\,AU, 
while those with slower rotation ($\beta_0 < 0.02$; '$\times$') 
in the later epoch ($\nc \simeq 10^{18}-10^{21}\cm$) with 
smaller separation $\simeq 0.01-1$\,AU.
These two fragmentation epochs have their origin in thermal evolution 
of primordial gas.
The lower-density fragmentation is a result of a
temperature drop owing to the three-body H$_2$ formation
at $\nc \simeq 10^{10}\cm$, while the higher-density one results from another temperature drop at $\nc \simeq 10^{17}\cm$, owing to the H$_2$ collision-induced cooling.
Recall that a quasi-rotation supported disk can continue to collapse for 
$\gamma <1$ while conserving angular momentum.
Thus even if collapse of the disk is about to halt, 
the disk can collapse further owing to the temperature drops.
Such disks stop their collapse and fragment when the temperature 
starts increasing again.
It is known that these two cooling mechanisms do not cause 
fragmentation of the core by simple thermal instability 
(Omukai \& Yoshii 2003; Ripamonti \& Abel 2004).
In our case, where a disk is present, 
these cooling mechanisms result in the fragmentation of the disk.

There is a caveat in our results on fragmentation epochs.
We adopted the equation of state derived from one-zone calculation.
For example, the temperature variation in one-zone models
tends to be exaggerated: the two temperature dips are deeper than 
in hydrodynamical calculations (e.g., Omukai \& Nishi 1998).
In addition, in hydrodynamical models, temperature in the envelope 
tends to be higher than that in the center for the same density 
\citep{whitehouse06,stamatellos07}.
Despite these possible differences, we expect that the fragmentation 
condition itself (eq. \ref{eq:beta0}) remains valid.
In deriving this condition (Sec. 6), we only assumed that 
once a quasi-rotation supported disk forms, it eventually fragments.
This is correct if the EOS has a sufficiently large density range 
with $\gamma >1$, which is indeed the case in the primordial gas.
In fact, in the context of present-day star formation, the same 
fragmentation criterion has been reported \citep{matsu03,machida05b}.
On the other hand, the fragmentation epoch and scale can be subject to 
change if EOS is altered.

Although the majority of cores are predicted to fragment before the stellar 
core formation, whether they finally evolve into bona fide binaries/multiples 
remains obscure.
The separation between fragments changes significantly 
during the mass accretion phase as a result of angular momentum influx to
the fragments.
An accreted gas in a later stage tends to have a larger specific angular momentum,
while spiral arms in a density pattern, which transfer   
angular momentum outward, might be invoked in such a phase. 
To answer whether fragments evolve into binaries/multiples or 
merge into a single star, we need to extend calculations 
over the entire accretion phase.

\begin{figure}
\begin{center}
\vspace{-1cm}
\includegraphics[width=130mm]{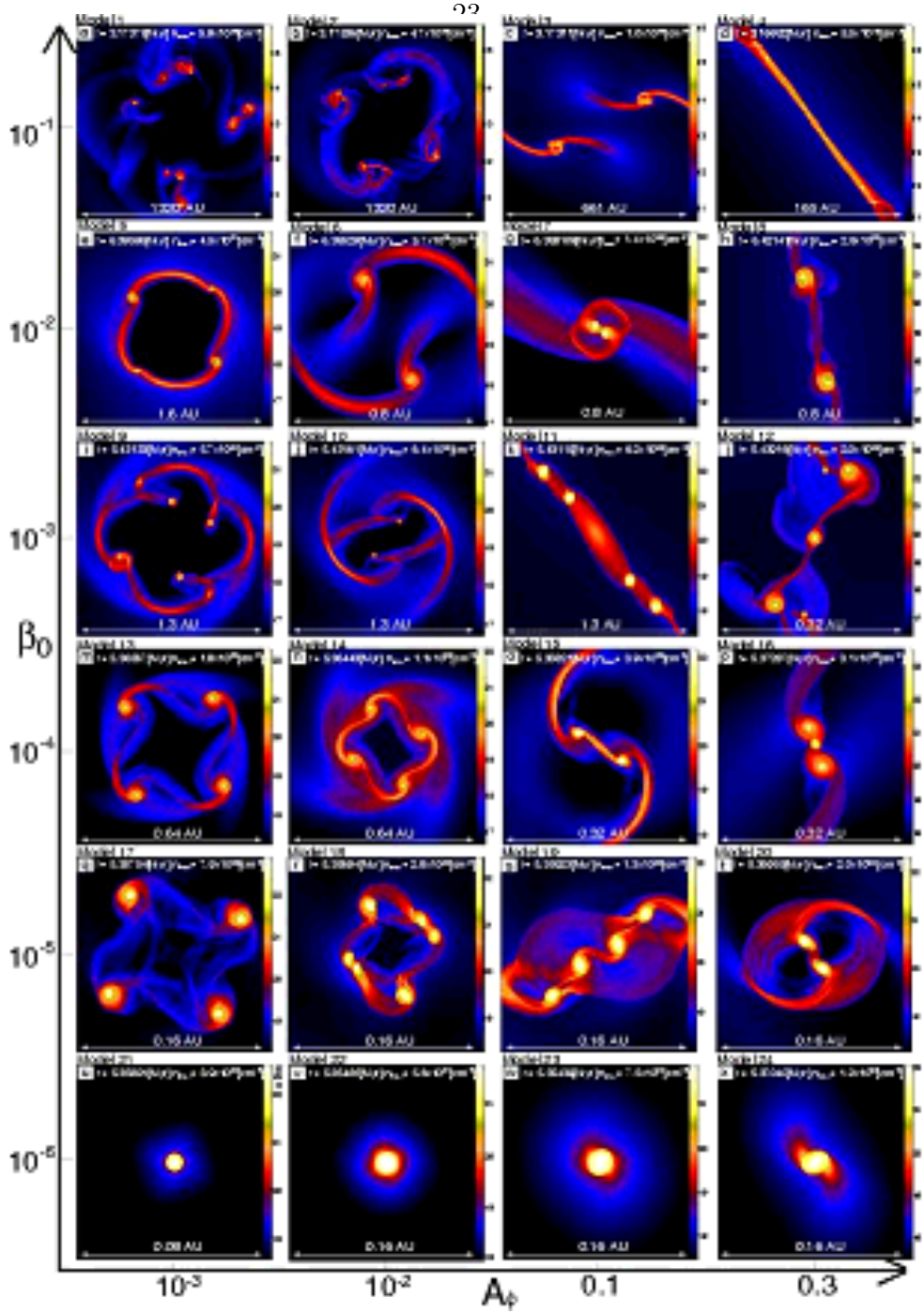}
\caption{The collapse outcome of primordial-gas cores with different
initial rotation parameter $\beta_0$ and 
amplitude of non-axisymmetric (bar-mode) perturbation $\ap$.
The final states in our calculation for models 1-24 in Table 1, 
where the cores are close to hydrostatic initially, 
i.e., $\alpha_0 \sim 1$ are presented.
Shown by false colors is the density distribution around the center 
on the plane perpendicular to the rotation axis.
The model numbers are denoted on the upper-left corners of the panels.
The name of the panel (a-x), the time elapsed since the beginning of 
the calculation $t$, the maximum number density attained $n_{\rm max}$ 
are given at the top, and the box scale is shown at the bottom of the panel.
}
\label{fig:final}
\end{center}
\end{figure}

\begin{figure}
\begin{center}
\includegraphics[width=140mm]{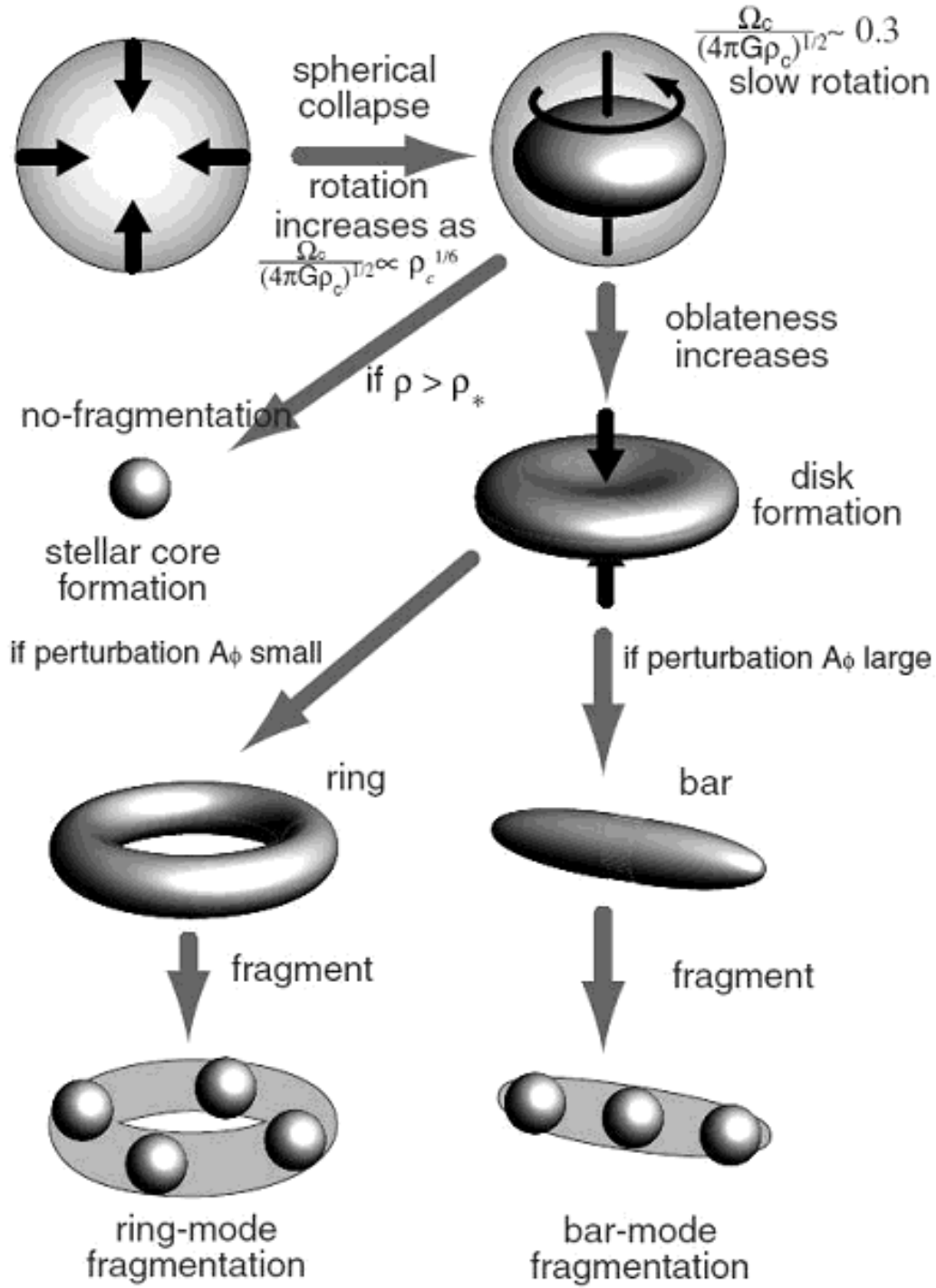}
\caption{A schematic for the evolution of rotating cores.
If the core reaches the stellar density $\rho_*$ before the disk formation, 
it does not fragment.
Otherwise, it fragments in the bar mode or ring mode
depending on the amplitude of the bar-mode perturbation 
at the time of disk formation. 
}
\label{fig:schematic}
\end{center}
\end{figure}

\begin{figure}
\begin{center}
\includegraphics[width=140mm]{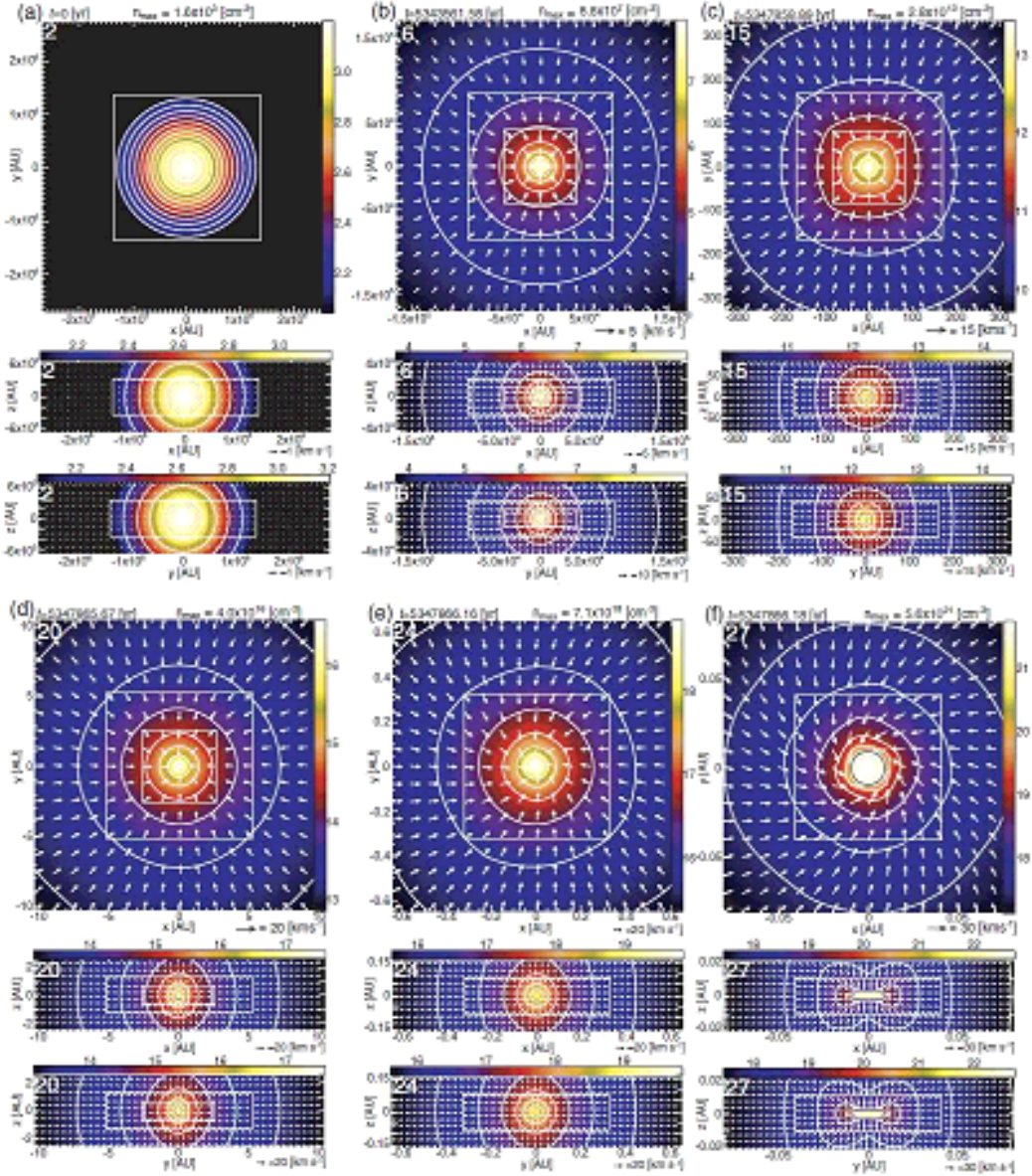}
\caption{
The density ({\it color scale and contour}) and velocity ({\it arrows})
distributions at six different epochs in model 22 
[($\alpha_0$, $\beta_0$, $\ap$) = (0.83, $10^{-6}$, $10^{-2}$)].
The upper, middle, and lower panels show those on $x-y$, $x-z$, 
and $y-z$ planes, respectively.
The elapsed time, maximum density (above) and arrow scale (below) 
are denoted above or below each panel.
The level of the grid is indicated in the upper left corner 
of each panel.
Solid squares in each panel are outer boundaries of subgrids.
The structure around the center is zoomed up from Panel {\it a} 
toward {\it f}.
The grid size of Panel {\it a} (the coarsest grid shown is $l=2$) 
is $6.7\times 10^6$ AU, 
while that of Panel {\it f} ($l=27$) is $0.2$ AU, 
$2^{25}$ times finer than in Panel {\it a}.
The black contour in Panel {\it f} denotes the constant density 
surface of $n=10^{20}\cm$, approximately corresponding to the surface of 
the stellar core.
}
\label{fig:model22}
\end{center}
\end{figure}

\begin{figure}
\begin{center}
\includegraphics[width=80mm]{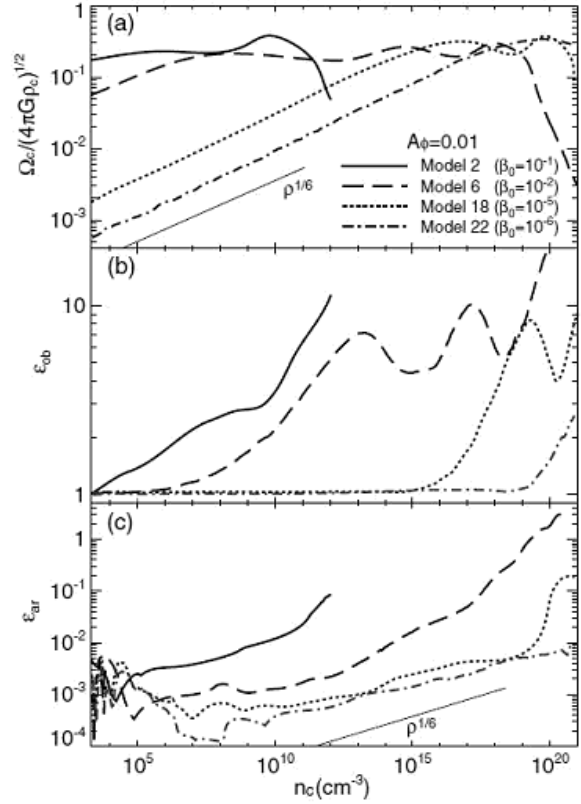}
\caption{Evolution of 
({\it a}) the central normalized angular velocity $\omega_{\rm c}$,
({\it b}) oblateness $\ob$ and 
({\it c}) axis ratio $\ar$ of the cores 
as a function of the central number density $\nc$.
Four cases (models 2, 6, 18 and 22 in the second column of 
Figure \ref{fig:final}) with the same initial non-axisymmetric 
perturbation $A_{\phi}=0.01$ but with different rotation parameters 
($\beta_0=10^{-6}-10^{-1}$) are shown.
The dash in Panel ({\it a}) shows the growth rate for the spherical collapse, 
while that in Panel ({\it c}) indicates the growth rate in the isothermal case
by the linear analysis.
}
\label{fig:5a}
\end{center}
\end{figure}

\begin{figure}
\begin{center}
\includegraphics[width=160mm]{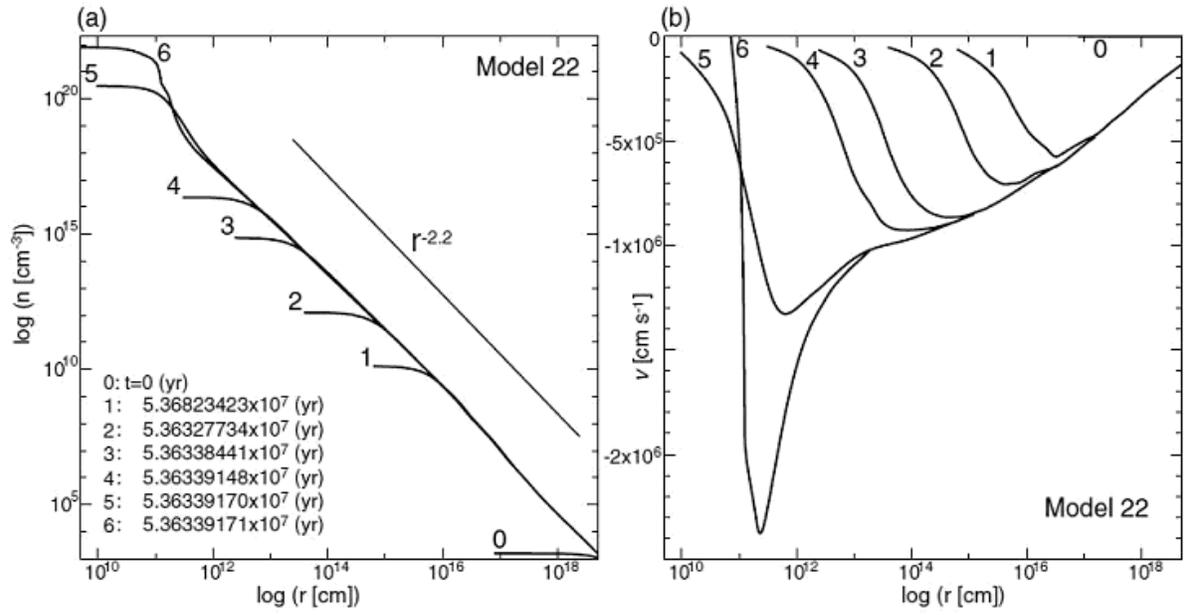}
\caption{
The radially averaged (a) number density and (b) velocity profiles 
at different epochs in model 22.
The time for each plot is indicated in panel {\it a}.
The dash in panel {\it a} shows the relation $n \propto r^{-2.2}$.
}
\label{fig:6a}
\end{center}
\end{figure}

\begin{figure}
\begin{center}
\includegraphics[width=160mm]{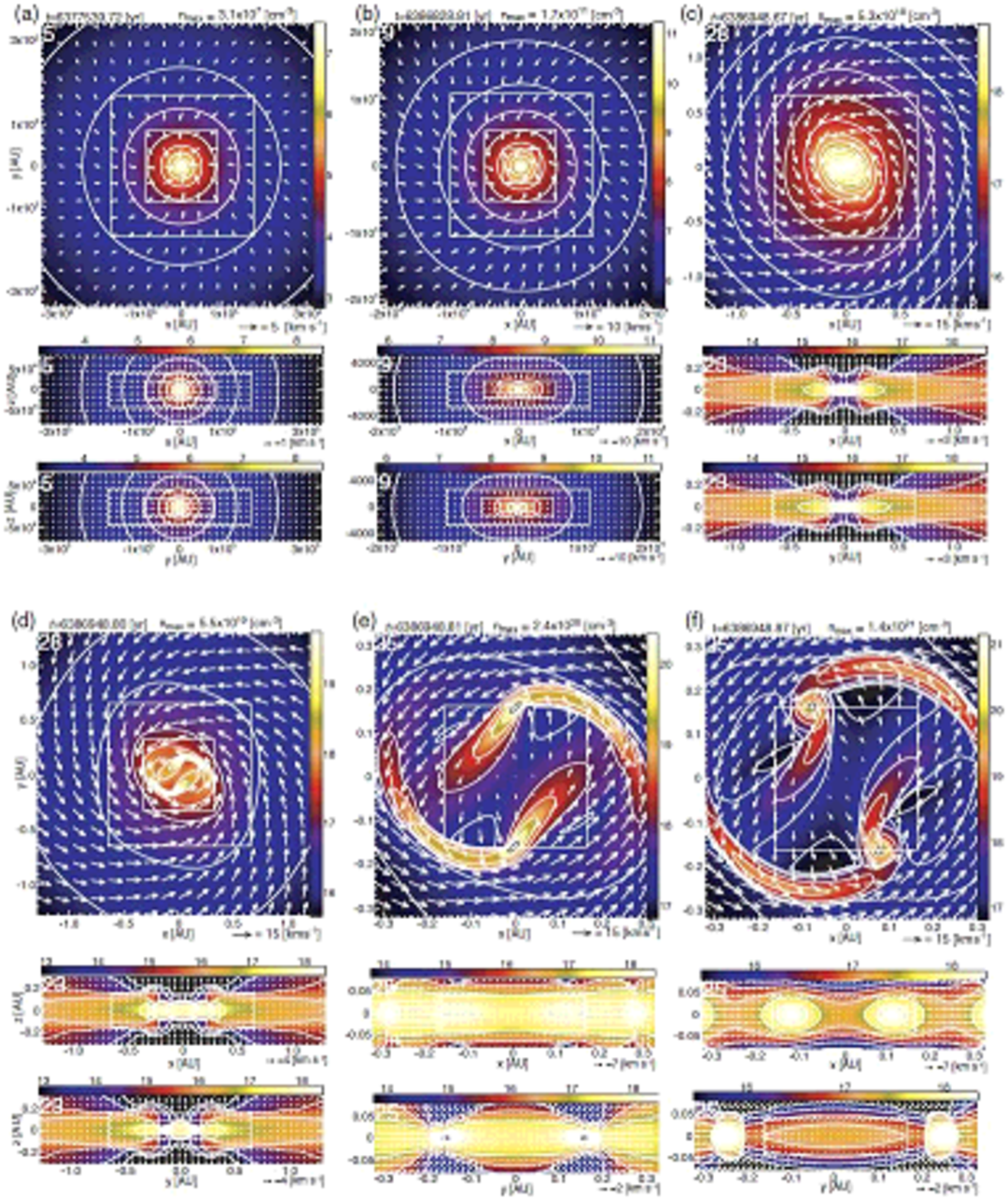}
\caption{
The same as Fig.~\ref{fig:model22} but for model 6 with 
($\alpha_0$, $\beta_0$, $\ap$) = (0.83, $10^{-2}$, 0.01).
}
\label{fig:7}
\end{center}
\end{figure}

\begin{figure}
\begin{center}
\includegraphics[width=160mm]{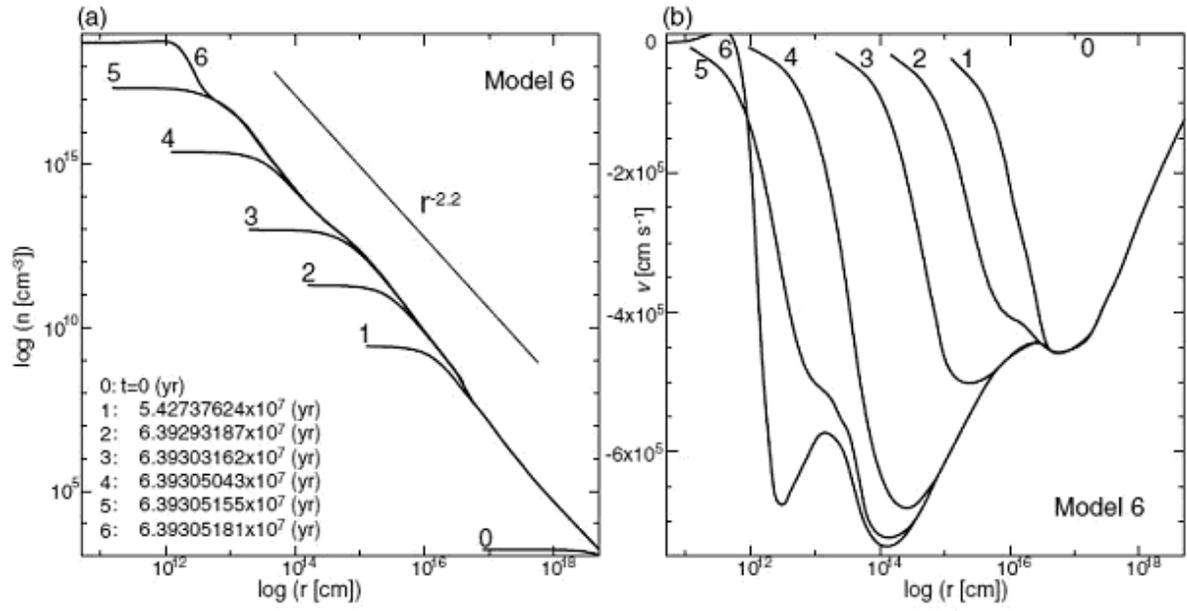}
\caption{
The same as Figure \ref{fig:6a} but for model 6.
}
\label{fig:6b}
\end{center}
\end{figure}

\begin{figure}
\begin{center}
\includegraphics[width=140mm]{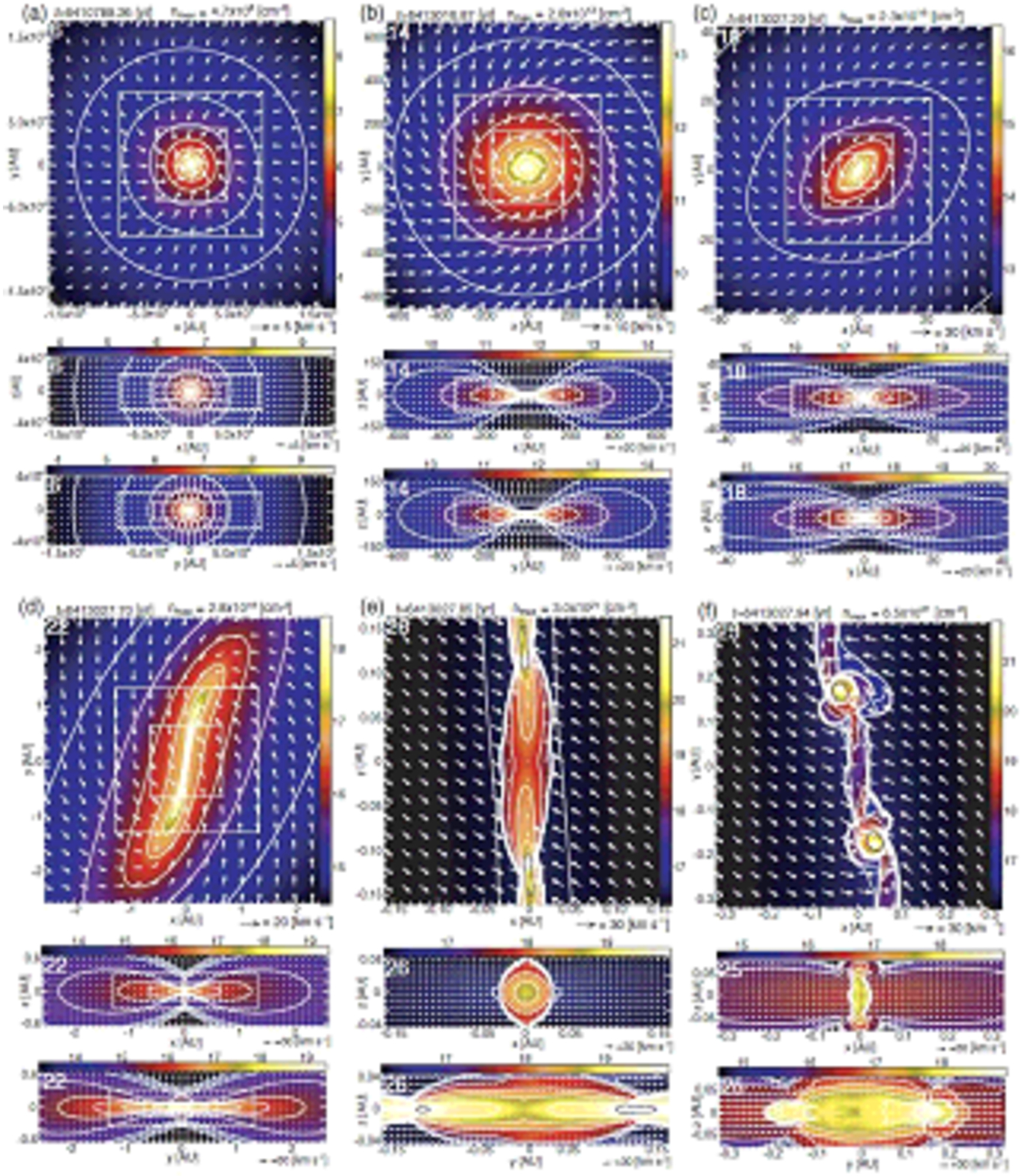}
\caption{
The same as Fig.~\ref{fig:model22} but for model 8 with 
($\alpha_0$, $\beta_0$, $\ap$) = (0.83, $10^{-2}$, 0.3).
}
\label{fig:8}
\end{center}
\end{figure}

\begin{figure}
\begin{center}
\includegraphics[width=80mm]{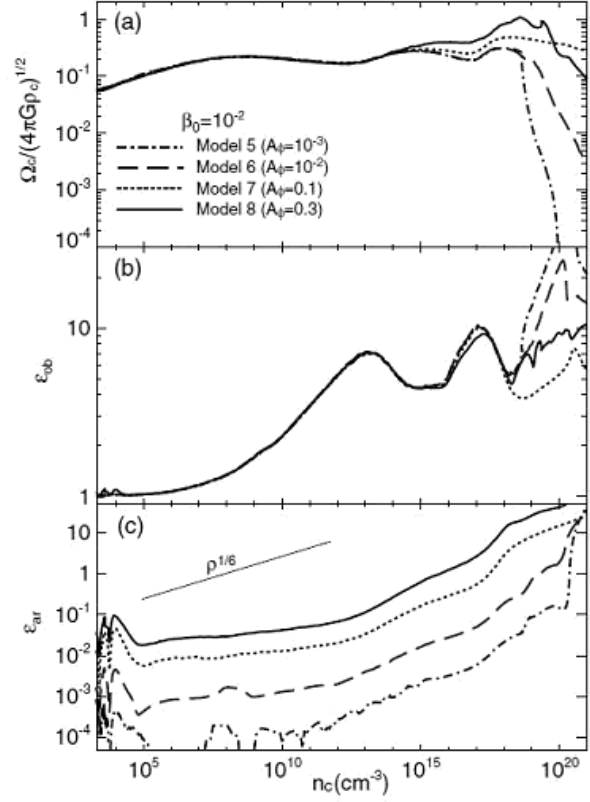}
\caption{
The same as Figure \ref{fig:5a} but for cases with same initial rotation 
parameter $\beta_0=10^{-2}$ with different amplitude of initial 
non-axisymmetric perturbation $A_{\phi}=0.01$
(models 5, 6, 7 and 8, in the second row of Figure \ref{fig:final}).
}
\label{fig:5b}
\end{center}
\end{figure}

\begin{figure}
\begin{center}
\includegraphics[width=140mm]{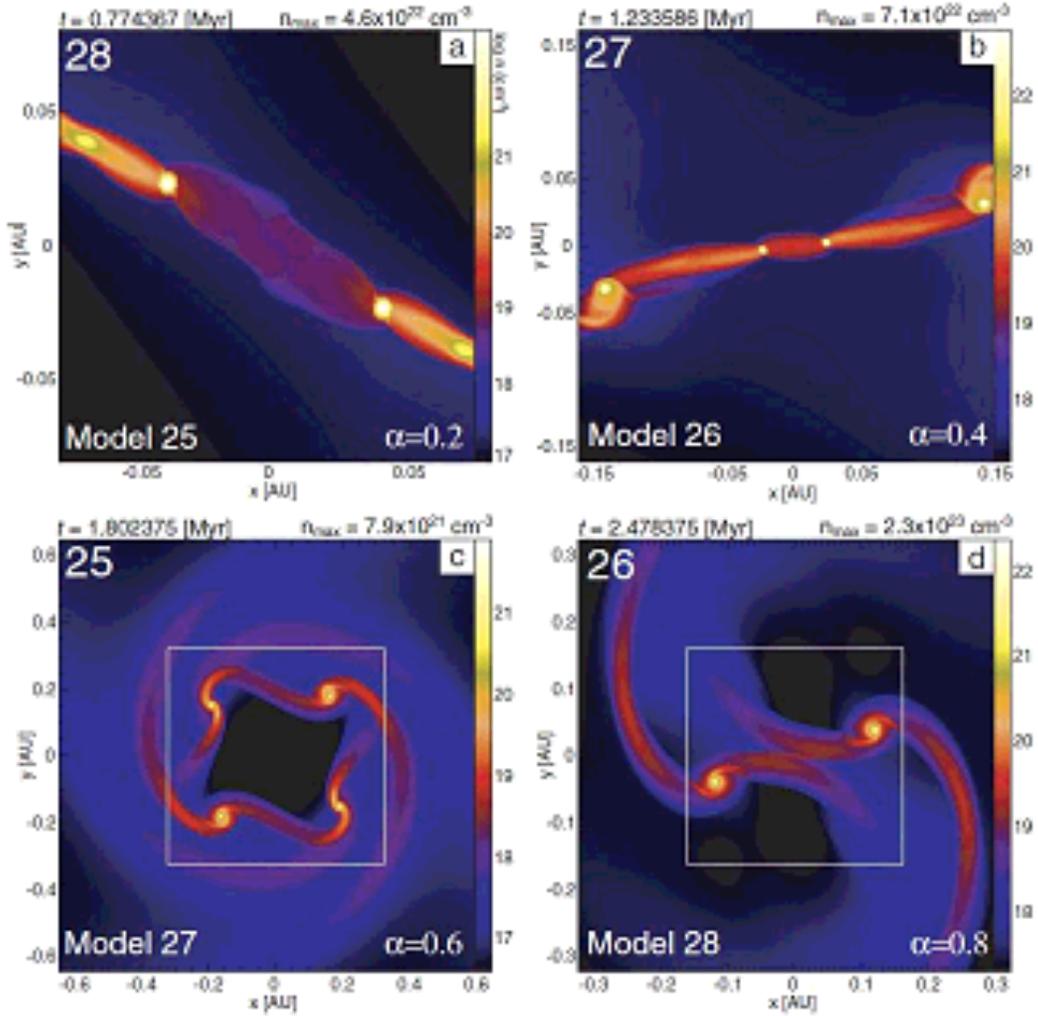}
\caption{
The collapse outcome of highly gravitationally unstable cores 
(models 25-28). 
The values of $\alpha_0$ are indicated in the bottom-left corner of 
the panels.
The legend is the same as in Figure \ref{fig:model22}.
}
\label{fig:9}
\end{center}
\end{figure}

\begin{figure}
\begin{center}
\includegraphics[width=120mm]{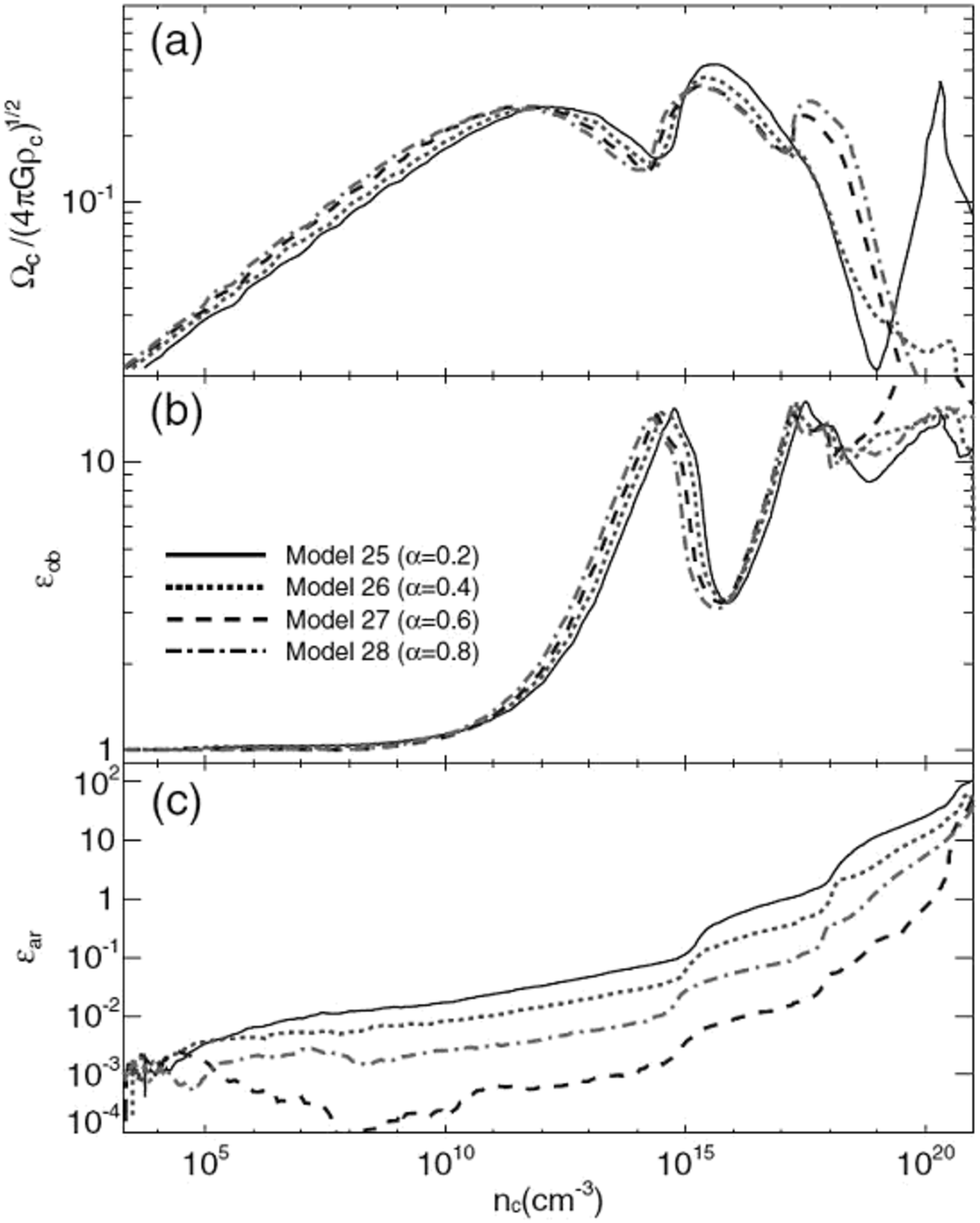}
\caption{
The same as Figure \ref{fig:5a} but for highly gravitationally unstable 
cores (models 25-28).
The values of $\alpha_0$ are indicated in panel (b).
}
\label{fig:10}
\end{center}
\end{figure}

\begin{figure}
\begin{center}
\includegraphics[width=130mm]{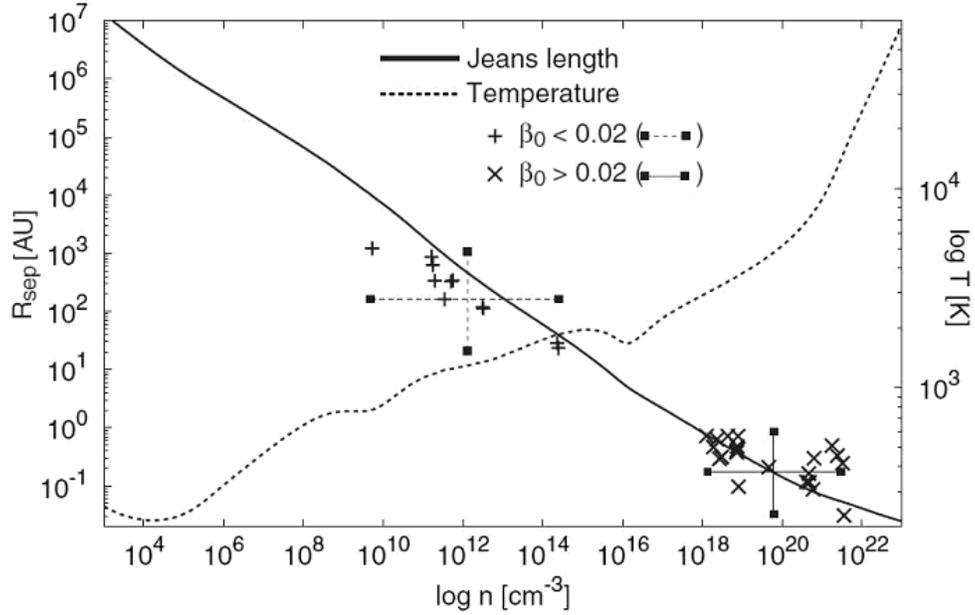}
\caption{
The separations between fragments against the density at fragmentation. 
We adopt the separation between furthermost fragments 
if more than two fragments appeared.
The evolution of temperature (right) and Jeans length (left) 
is also illustrated 
by the dotted and solid lines, respectively.
The results of all the models where fragmentation is observed are plotted.
Those with the rotation parameter $\beta_0>0.02$ ($< 0.02$) are shown by 
symbols `+',  (`$\times$', respectively).
The large dotted (solid) cross represents the range of the fragment separation 
and fragmentation epoch for cases with $\beta_0 > 0.02$ ($< 0.02$, 
respectively).
}
\label{fig:12}
\end{center}
\end{figure}

When more than two fragments appear 
(e.g., models 1 and 2 in Fig.~\ref{fig:final}), 
some might be ejected from the parental core 
as a result of three-body interactions \citep{bate02,bate03}.
Ejected objects stop gaining mass by accretion thereafter.
By this mechanism, a spectrum in mass of metal-free stars 
might be formed.
For example, if an ejection event takes place in the very early accretion phase, 
the ejected object can eventually evolve into a metal-free brown dwarf.
Or if a stellar core with its envelope of $\sim 1\msun$ is ejected, 
a low-mass metal-free star is the outcome.
Considering the origin of the two hyper metal-poor ([Fe/H]$<-5$) 
stars (Christlieb et al. 2001; Frebel et al. 2005), 
\citet{suda04} showed that their abundance patterns 
can be explained by nucleosynthesis and mass transfer 
in first-generation low-mass binary stars.
In our scenario, such a system can be formed naturally.
In some models (e.g., see Fig.~\ref{fig:final}{\it a} and {\it b}),
multiple pairs of binaries appear.
If one such pair is ejected from the parental core, 
it would result in a low-mass metal-free binary system.

With neither ejection nor merging to a single star, our binary systems 
of fragments are expected to become very massive star binaries.
Such systems are considered to be important sources of gravitational waves 
(Belczynski et al. 2004).
Also they can be gamma-ray burst progenitors if formed in close binaries 
\citep{bromm06,yoon06}.

In this study, we did not include magnetic effects.
According to recent studies, however, a low level of magnetic field could be 
present even in the early universe \citep{langer03,ichiki06}.
If an initial primordial-gas core has $B>10^{-13.5}$\,G of the magnetic field, 
a protostellar jet is launched \citep{machida06c}, and this jet significantly 
affects the mass accretion process.
Even with a lower level, a magnetic field can suppress fragmentation 
of the core, 
by transferring angular momentum via magnetic braking 
\citep[e.g.,][]{machida07, price07}.
So far those effects on primordial star formation have been largely overlooked. 
Further studies along this line are clearly needed in future.

\acknowledgments
We have greatly benefited from discussion with ~H. Susa, ~T. Tsuribe, 
and ~K. Saigo.
We also thank T. Hanawa for contribution to the nested grid code.
Numerical computations were carried out on VPP5000 at Center for Computational Astrophysics, CfCA, of National Astronomical Observatory of Japan.
This work is supported by the Grant-in-Aid for the 21st Century COE "Center for Diversity and Universality in Physics" from the Ministry of Education, Culture, Sports, Science and Technology (MEXT) of Japan, and partially supported by 
the Grants-in-Aid from MEXT (15740118, 16077202, 18740104, 18740113).

\begin{table}   
\vspace{-1cm}
\caption{Model parameters and calculation results}
\label{table:init}
\begin{center}
\begin{tabular}{c|ccc|ccccccccccccccccc}
\hline
Model & $\beta_0 $ & A$_{\rm \phi}$ & $\alpha_0$ ($f$) & $\nf$ {\scriptsize ($\cm$)}$^a$& NF$^b$ & Sep.$^c$ & Mode$^d$ \\
\hline
1 & 0.1      & $10^{-3}$& 0.7 (1.20) & 1.62$\times10^{11}$ & 8 & 870 & R   \\
2 & 0.1      & 0.01     & 0.7 (1.20) & 1.76$\times10^{11}$ & 6 & 625 & R   \\
3 & 0.1      & $0.1    $& 0.7 (1.20) & 1.95$\times10^{11}$ & 2 & 335 & R   \\
4 & 0.1      & $0.3    $& 0.7 (1.20) & 3.47$\times10^{11}$ & 8 & 163 & B   \\
\hline
5 & 0.01& $10^{-3}$& 0.83 (1.01) & 4.34$\times10^{18}$ & 4 & 0.722& R   \\
6 & 0.01& 0.01     & 0.83 (1.01)& 7.08$\times10^{18}$ & 2 & 0.385 & R   \\
7 & 0.01& $0.1    $& 0.83 (1.01)& 7.25$\times10^{18}$ & 2 & 0.377 & B   \\
8 & 0.01& $0.3    $& 0.83 (1.01)& 1.89$\times10^{19}$ & 2 & 0.463 & B   \\
\hline
9 & $10^{-3}$& $10^{-3}$& 0.83 (1.01) & 2.31$\times10^{18}$ & 6 & 0.632 & R  \\
10& $10^{-3}$&0.01      & 0.83 (1.01)& 2.66$\times10^{18}$ & 2 & 0.292 & R   \\
11& $10^{-3}$& $0.1    $& 0.83 (1.01)& 3.03$\times10^{18}$ & 2 & 0.314 & R   \\
12& $10^{-3}$& $0.3    $& 0.83 (1.01)& 7.43$\times10^{19}$ & 5 & 0.476 & B   \\
\hline
13& $10^{-4}$& $10^{-3}$& 0.83 (1.01)& 7.54$\times10^{18}$ & 4 & 0.419 & R   \\
14& $10^{-4}$& 0.01     & 0.83 (1.01)& 7.78$\times10^{18}$ & 4 & 0.714 & R   \\
15& $10^{-4}$& $0.1    $& 0.83 (1.01)& 7.99$\times10^{18}$ & 2 & 0.097 & B   \\
16& $10^{-4}$& $0.3    $& 0.83 (1.01)& 3.25$\times10^{21}$ & 3 & 0.245 & R   \\
\hline
17& $10^{-5}$& $10^{-3}$& 0.83 (1.01)& 4.03$\times10^{20}$ & 4 & 0.117 & R   \\
18& $10^{-5}$& 0.01     & 0.83 (1.01)& 5.92$\times10^{20}$ & 6 & 0.087 & R   \\
19& $10^{-5}$& $0.1    $& 0.83 (1.01)& 4.77$\times10^{20}$ & 4 & 0.115 & R   \\
20& $10^{-5}$& $0.3    $& 0.83 (1.01)& 3.53$\times10^{21}$ & 2 & 0.031 & R   \\
\hline
21& $10^{-6}$& $10^{-3}$& 0.83 (1.01)& --- & 1 & --- & ---   \\
22& $10^{-6}$& 0.0     1& 0.83 (1.01)& --- & 1 & --- & ---   \\
23& $10^{-6}$& $0.1    $& 0.83 (1.01)& --- & 1 & --- & ---   \\
24& $10^{-6}$& $0.3    $& 0.83 (1.01)& --- & 1 & --- & ---   \\
\hline
25& $10^{-3}$& 0.01& 0.2 (4.19)& 4.62$\times10^{20}$ & 4 & 0.163 & B  \\
26& $10^{-3}$& 0.01& 0.4 (2.10)& 6.24$\times10^{20}$ & 4 & 0.298 & B  \\
27& $10^{-3}$& 0.01& 0.6 (1.40)& 1.76$\times10^{21}$ & 4 & 0.492 & R  \\
28& $10^{-3}$& 0.01& 0.8 (1.05)& 2.41$\times10^{21}$ & 2 & 0.330 & R  \\
\hline
29& 0.02& 0.01& 0.6 (1.40)& 1.26$\times10^{18}$ & 2 & 0.720 & R  \\
30& 0.04& 0.01& 0.6 (1.40)& 2.46$\times10^{14}$ & 4 & 23.4 & R   \\
31& 0.06& 0.01& 0.6 (1.40)& 3.10$\times10^{12}$ & 4 & 119 & R    \\
32& 0.08& 0.01& 0.6 (1.40)& 5.03$\times10^{11}$ & 4 & 328 & R    \\
33& 0.2 & 0.01& 0.6 (1.40)& 5.23$\times10^{9}$ & --- &1.22$\times10^{3}$ & R \\
\hline
34& 0.02& 0.1& 0.6 (1.40)& 7.16$\times10^{19}$ & 2 & 0.209 & B  \\
35& 0.04& 0.1& 0.6 (1.40)& 2.29$\times10^{14}$ & 4 & 28.8 & R   \\
36& 0.06& 0.1& 0.6 (1.40)& 3.21$\times10^{12}$ & 4 & 113 & R    \\
37& 0.08& 0.1& 0.6 (1.40)& 5.51$\times10^{11}$ & 4 & 344 & R    \\
38& 0.2&  0.1& 0.6 (1.40)& 5.38$\times10^{9}$ & --- & 1.21$\times10^{3}$ & R &  \\
\hline
\end{tabular}
\\ 
$^a$ number density at fragmentation epoch, $^b$ number of fragments, $^c$ the separation between furthermost fragments, $^d$ fragmentation mode (R: ring, B: bar).
\end{center}
\end{table}

\clearpage
\clearpage
\clearpage
\clearpage
\clearpage
\clearpage
\clearpage
\clearpage
\clearpage
\clearpage
\clearpage
\clearpage
\clearpage
\end{document}